\documentclass[%
 aip,jcp,
 amsmath,amssymb,
 reprint,twocolumn,%
]{revtex4-1}
\usepackage[english]{babel}
\usepackage {CJK}
\usepackage{hyperref}
\usepackage{graphicx}
\usepackage{dcolumn}
\usepackage{bm}

\usepackage[utf8]{inputenc}
\usepackage[T1]{fontenc}
\usepackage{mathptmx}
\usepackage{amssymb}
\usepackage{color}
\begin{document}

\preprint{AIP/123-QED}

\title[]{Deep Learning for UV Absorption Spectra with SchNarc: First Steps Towards Transferability in Chemical Compound Space}

\author{J. Westermayr}
\affiliation{%
University of Vienna, Faculty of Chemistry, Institute of Theoretical Chemistry, W\"ahringer Str. 17, 1090 Vienna, Austria.}
\author{P. Marquetand}
 \email{philipp.marquetand@univie.ac.at}
 \affiliation{%
University of Vienna, Faculty of Chemistry, Institute of Theoretical Chemistry, W\"ahringer Str. 17, 1090 Vienna, Austria.}
\affiliation{
Vienna Research Platform on Accelerating Photoreaction Discovery, University of Vienna \\ W\"ahringer Str. 17, 1090 Vienna, Austria.
}
\affiliation{University of Vienna, Faculty of Chemistry, Data Science @ Uni Vienna, W\"ahringer Str. 29, 1090 Vienna, Austria.
}%

\date{\today}

\begin{abstract}
Machine learning (ML) has shown to advance the research field of quantum chemistry in almost any possible direction and has recently also entered the excited states to investigate the multifaceted photochemistry of molecules. In this paper, we pursue two goals: i) We show how ML can be used to model permanent dipole moments for excited states and transition dipole moments by adapting the charge model of [Chem. Sci., 2017, 8, 6924-6935], which was originally proposed for the permanent dipole moment vector of the electronic ground state. ii) We investigate the transferability of our excited-state ML models in chemical space, i.e., whether an ML model can predict properties of molecules that it has never been trained on and whether it can learn the different excited states of two molecules simultaneously. To this aim, we employ and extend our previously reported SchNarc approach for excited-state ML. We calculate UV absorption spectra from excited-state energies and transition dipole moments as well as electrostatic potentials from latent charges inferred by the ML model of the permanent dipole moment vectors. We train our ML models on CH$_2$NH$_2^+$ and C$_2$H$_4$, while predictions are carried out for these molecules and additionally for CHNH$_2$, CH$_2$NH, and C$_2$H$_5^+$. The results indicate that transferability is possible for the excited states.  
\end{abstract}
\maketitle

\section{\label{sec:Introduction}Introduction}
Photosynthesis,~\cite{Cerullo2002S,Romero2017N} the ability of beings to see, photorelaxation of e.g. DNA and proteins to prevent them from photodamage\cite{Schultz2004S,Schreier2007S,Rauer2016JACS} are fascinating examples of the importance of light-matter interactions for our daily lives. Another marvelous aspect are the colors of every thing and every being, which are related to the absorption of a part of the incident lights spectrum. In order to get a deeper understanding of these phenomena and to find out about the possibility of a molecule to be excited by light, answers to the following questions have to be provided: At which wavelengths can a molecule absorb electromagnetic radiation? How much of these wavelengths is absorbed? Can this absorption be used to identify a molecule?

In order to answer such questions, experiments or quantum chemical calculations are usually carried out. Assuming the resonance condition, i.e., the equivalence of the energy of one or more photons of the incident light with the energy gap between two electronic states, single- or multiphoton excitations can take place to one or more excited states if an oscillating dipole is induced.~\cite{He2008CR,Marquetand2014,Tagliamonti2016PRA}
The oscillator strength,$f^{osc}_{ij}$, between electronic state i and j, is related to the transition dipole moment, $\mu_{ij}$, of the respective electronic states as well as the energy difference, $\Delta E_{ij}$, between them~\cite{Hilborn1982AJP} and is given in a.u.: $f^{osc}_{ij}=\frac{2}{3} \Delta E_{ij}\mid\mu_{ij}\mid^2$.
The larger the oscillation strength, the more likely a transition takes place.

Corresponding experiments often lack the possibility to distinguish and characterize the different electronic states and rely on theoretical simulations to identify the states and provide detailed insights on their characters.  However, these calculations are limited by the high costs for solving the underlying quantum chemical equations. 
Especially the excited states necessitate highly accurate quantum chemical methods, whose computational costs scale unfavourably with the number of electronic states and atoms considered in the calculations.~\cite{Gonzalez2020,Nelson2020CR} Further, sampling of many different molecular configurations followed by statistical averaging is often required in order to accurately reproduce the shape of experimentally obtained spectra. The many calculations, which are needed to obtain accurate results seriously limit the calculations.

A solution to the aforementioned problems can be obtained with (atomistic) machine learning (ML) models, which have shown to be extremely powerful for the electronic ground state to provide ML potentials for energies or dipole moments, see e.g. refs ~\citenum{Bartok2010PRL,Li2015PRL,Rupp2015JPCL,Behler2016JCP,Artrith2017PRB,Gastegger2017CS,Deringer2017PRB,Botu2017JPCC,Smith2017CS,Zong2018npjCM,Chmiela2018NC,Imbalzano2018JCP,Zhang2018PRL,Chan2019JPCC,Wang2019JCTC,Gerrits2019JPCL,Chen2020JPCA_Dynamics,Netzloff2006JCP,Bettens1999JCP,Zubatyuk2019SA}. ML force fields exist~\cite{Christensen2020JCP,Schuett2018JCP,Rudorff2020PCCP,Zeni2019APX,Jinnouchi2019PRB,Botu2017JPCC,Chen2020JPCA_Dynamics,Behler2017ACIE,Behler2016JCP,Behler2007PRL,Artrith2011PRB,Deringer2017PRB} and also the transferability of properties has been demonstrated.~\cite{Montavon2013NJP,Hansen2015JPCL,Christensen2019JCP,Christensen2020JCP,Tawfik2019ATS,Christensen2019CHIMIA,Ghosh2019AS}
The main advantages of ML models is that they can sample a huge number of molecular configurations with the accuracy of the underlying quantum chemical calculations at only a fraction of the original costs.~\cite{Behler2017ACIE,Schuett2020}

Recently, the interest to advance also the research field of photochemistry and to tackle the excited states with ML has increased.~\cite{Westermayr2020MLST_Perspective,Westermayr2020arXiv_review,Dral2020} The fitting of molecule-specific potential energy surfaces (PESs) and coupling values~\cite{Carbogno2010PRB,Hu2018JPCL,Dral2018JPCL,Chen2018JPCL,Xie2018JCP,Guan2019PCCP,Westermayr2019CS,Guan2020JCTC,Richings2019JCTC,Guan2019JCP,Wang2019JPCA,Richings2018JCP,Richings2017CPL,Westermayr2020MLST,Westermayr2020JPCL,Westermayr2020MLST,Guan2020JPCL} or dipole moments as single values~\cite{Westermayr2019CS,Westermayr2020MLST,Guan2020JCTC} has been demonstrated up to date and energy gaps, HOMO-LUMO gaps as well as oscillator strengths have been fitted.~\cite{Xue2020chemrxiv,Ramakrishnan2015JCP,Ghosh2019AS,Pronobis2018EPJB,Schuett2019NC,Ye2019PNAS} The novel proposed ML models are mostly based on many configurations of a single molecule.~\cite{Dral2018JPCL,Chen2018JPCL,Westermayr2019CS,Westermayr2020JPCL,Westermayr2020MLST,Westermayr2020MLST_Perspective,Zhang2020arXiv,Guan2020JCTC,Guan2020JPCL,Jiang2016IRPC,Xie2018JCP,Richings2017CPL,Richings2020JCP,Richings2019FD} Only a few ML models treat different molecules in their energetic equilibrium structure, which is mapped to a single output, e.g. the oscillator strength.~\cite{Ramakrishnan2015CHIMIA} Yet is unclear, whether such a universal ML force field as it exists for the electronic ground state is feasible also for the excited states. The description of many molecular systems with one ML model further requires the construction of excited-state properties from atomistic contributions, but most ML models targeting the excited states employ molecule-wise descriptors and some studies suggest molecular descriptors to be superior to atom-wise descriptors for the excited states.~\cite{Westermayr2020MLST_Perspective,Westermayr2020arXiv_review,Pronobis2018EPJB}

Another limitation of many existing ML models for the computation of absorption spectra is that they fit the oscillator strengths rather than the excited-state energies and transition dipole moments. The fitting of the latter properties is beneficial as they can be used e.g. for the computation of photodynamics, ML/MM (ML/molecular mechanics) schemes~\cite{Lahey2020CS} similar to QM/MM (quantum mechamics/MM) schemes~\cite{Menger2017JCTC} or the investigation of explicit light interaction\cite{Richter2011JCTC,Marquetand2011FD} -- to name only a few applications. 

Transition dipole moments and permanent dipole moments can be computed by applying the dipole moment operator as implemented in many electronic structure programs. Permanent dipole moments can also be constructed from atomic charges using the point charge model (eq. \ref{eq:mui}). By having access to the atomic charges of a molecule, not only the dipole moment vector can be computed, but also the charge fluctuations within dynamics or different reaction coordinates can be investigated and electrostatic potentials can be computed~\cite{Tomasi1996,Gastegger2020}. The atomic charges of the excited states can further be used to construct approximated excited-state force fields~\cite{Heid2016JCP} or can be used to investigate how the charge distribution changes due to light excitation. Although atomic charges are considered one of the most intuitive chemical concepts, they cannot be obtained directly by solving the Schr\"odinger equation.~\cite{Szabo2012} Subsequent analysis of the charge distribution in a molecule is highly dependent on the underlying partitioning scheme applied.~\cite{Wiberg1993JCC}

Dipole models based on ML~\cite{Ramakrishnan2014SD,Artrith2011PRB,Huang2016JCP,Gastegger2017CS,Yao2018CS,Schuett2018JCP,Nebgen2018JCTC,Sifain2018JPCL,Schuett2019JCTC,Willat2019JCP,Veit2020JCP}
can provide access to the density or latent partial charges while being based on the underlying electronic structure theory. For instance, the latent charges obtained from the dipole moment ML model reported in ref.~\citenum{Gastegger2017CS}, which never learn atomic charges directly, show good agreement with common charge models (CHELPG~\cite{Breneman1990JCC} and Hirshfeld~\cite{Hirshfeld1977TCA}), which are considered to be more reliable than for example Mulliken~\cite{Mulliken1955JCP} charges.~\cite{Gastegger2020} 
They have been used to plot electrostatic potentials and assess the changes of atomic charges with respect to molecular geometries for the electronic ground state in Refs.~\citenum{Gastegger2020,phdmichael}. Electrostatic potentials are further interesting~\cite{Tomasi1996} to interpret noncovalent interactions~\cite{Orozco1996}, for Quantitative Structure-Activity Relationship~\cite{Breneman1996} or for force fields.\cite{Koner2020JCP} 

Unfortunately, especially the fitting of transition dipole moments is challenging, as the sign of properties resulting from two different electronic states is arbitrary due to the arbitrary phase of the wave function,~\cite{Akimov2018JPCL,Westermayr2019CS} and because rotational covariance has to be preserved for vectorial properties. 
To the best of our knowledge, only one study exists,\cite{Zhang2020arXiv} in which phase corrected transition dipole moments were treated in a rotationally covariant way and a single-state fashion~\cite{Westermayr2020MLST} with ML. The trained ML models were used to fit model Hamiltonians for subsequent prediction of UV spectra. Yet an ML model that can describe many different PESs, forces, and dipole moment vectors simultaneously for the prediction of UV spectra does not exist.

In this work, we adapt the aforementioned ground-state charge model to describe the permanent dipole moments of the excited states and in addition, we extend it to model the transition dipole moments in a rotationally covariant way. 
To this aim, we use the SchNarc deep learning approach, originally developed for photodynamics simulations, to additionally enable the computation of UV spectra. By doing so, we extend the SchNarc approach enabling a simultaneous modelling of permanent and transition dipole moment vectors of an arbitrary number of electronic states in addition to a manifold of excited-state potentials, forces, and couplings thereof.
The methylenimonium cation, CH$_2$NH$_2^+$, and the isoelectronic molecule ethylene, C$_2$H$_4$, are used as model systems to asses the accuracy of ML-fitted transition dipole moments and latent partial charges by computation of UV spectra and electrostatic potentials. 

In addition, we aim to evaluate the possibility of training one ML model on a set of molecular conformations of CH$_2$NH$_2^+$ and C$_2$H$_4$, i.e. a multi-molecule model. The performance of this model is assessed by comparison to the single-molecule ML models. As SchNarc constructs energies and dipole moments from atomic contributions, the transferability of this model toward other molecules not included in the training set is evaluated. Thus, in addition to CH$_2$NH$_2^+$ and  C$_2$H$_4$, the molecules CH$_2$NH, CHNH$_2$, and C$_2$H$_5^+$ are described, which are not included in the training set and have never been seen by the ML model.

\section{Theory}
Recently, we reported the SchNarc approach for efficient photodynamics simulations with ML-fitted PESs, derivatives, and couplings.~\cite{Westermayr2020JPCL} In this work, we extend this deep learning model to fit permanent and transition dipole moments in a rotationally covariant way of an arbitrary number of states and pairs of states, respectively. 

In order to train an ML model on the different excited-state PESs and properties simultaneously, a training set has to be provided that consists of molecular conformations on the one hand and the corresponding PESs, forces, and excited-state properties on the other hand. The molecular geometries are automatically transformed into molecular descriptors by SchNet~\cite{Schuett2018JCP} and are intrinsic to the network architecture, which further relates these tailored molecular descriptors to the excited-states properties in an end-to-end fashion.~\cite{Schuett2019JCTC,Westermayr2020JPCL} 
The loss function, $L_{SchNarc}$, which is used to monitor the error on the different properties during training includes permanent and transition dipole moments, summarized in the term $\mu$, in addition to energies, forces, and different types of couplings:
\begin{equation}
\begin{array}{ll}\label{eq:loss}
       L_{SchNarc} =& t_E \cdot L_E
       + t_F \cdot L_F \\
       &+ t_{SOC} \cdot L_{SOC} 
       + t_{NAC} \cdot L_{NAC}+t_{\mu} \cdot L_{\mu}
       \end{array}
\end{equation}
The trade-offs between the errors of the different properties are labelled with the letter, $t$, and the error of each property with $L$. The subscripts E, F, $\mu$, SOC, and NAC denote energies, forces, permanent and transition dipole moments, spin-orbit couplings, and nonadiabatic couplings, respectively. The couplings are not accounted for in this work, as they spin-orbit couplings arise between states of different spin multiplicity~\cite{Penfold2018CR} and nonadiabatic couplings can be approximated from Hessians of the PESs.~\cite{Westermayr2020JPCL}

While the energies and forces can be monitored using the mean squared error (MSE) between predicted properties by the ML model (denoted with the superscript "ML") and the quantum chemical reference values (denoted as "QC"),
\begin{equation}\label{eq:l2}
\begin{array}{l}
L_E = \mid\mid E^{QC} - E^{ML} \mid\mid^2, ~\text{and}~ \\
      L_F= \mid\mid F^{QC}-F^{ML}\mid\mid^2 \\,
\end{array}
\end{equation}
a phase-less loss function has to be applied for coupling values and transition dipole moments unless they are phase corrected.~\cite{Westermayr2019CS} Different variants of such a phase-free training algorithm have been proposed by us, which depend on the type of calculation and can be found in detail in Ref.~\citenum{Westermayr2020JPCL}. SchNarc automatically determines the most suitable phase-free training process, which is in its simplest form the minimum function of the MSEs assuming once a negative and once a positive sign of a coupling value or dipole moment, respectively. The minimum function can be used when only one excited-state property with arbitrary signs is treated, which is the case here:
\begin{equation}
\begin{array}{l}
     L_\mu= min\left( \left\{\varepsilon_\mu^+, \varepsilon_\mu^- \right\} \right)
\end{array}
\end{equation}
with 
\begin{equation}
    \varepsilon_\mu^\pm = \mid\mid \mu^{QC} \pm \mu^{ML} \mid\mid.
\end{equation}
The dipole moments are treated as vectorial properties and thus the signs within a vector are conserved. As permanent dipole moment vectors are described together with transition dipole moment vectors, they are also trained in a phase-free manner. As a consequence, they are only defined up to an arbitrary sign, which can lead to permanent dipole moment vectors pointing into the wrong direction. Hence they have to be adjusted for a reference molecular geometry when making predictions. A more detailed discussion can be found in the supporting information (SI).~\cite{Westermayr2020JPCL}

The model for permanent and transition dipole moments used here is based on the charge model of Ref.~\citenum{Gastegger2017CS}: Since the ML model is an atomistic one, atomic contributions to the molecular dipole moment can be automatically obtained. These atomic contributions are taken as latent atomic charges, i.e., they have to be multiplied by the distance, $r_a^{CM}$, of the atoms, $a$, to the center of mass of the molecule and are then summed up, before feeding the resulting dipole moment into the loss function. In the same way as the permanent dipole moment of the electronic ground state, SchNarc fits the permanent dipole moment of arbitrary states, $\mu_i$, and transition dipole moments, $\mu_{ij}$, between different electronic states according to equations \ref{eq:mui} and \ref{eq:muij}, respectively.
\begin{equation}\label{eq:mui}
    \mu_{i} = \sum_a^{N_a} q_{i,a} r^{CM}_a
\end{equation}
\begin{equation}\label{eq:muij}
    \mu_{ij} = \sum_a^{N_a} q_{ij,a} r^{CM}_a
\end{equation}
Note that also here the atomic charges are latent variables and the "atomic transition charges" between two different states used to obtain transition dipole moments do not have a direct physical meaning. However, these charges are the quantities that are used in the predictions. They are then multiplied with $r_a^{CM}$ and, in this way, allow for rotational covariance of the transition dipole moment vectors.

\section{Models and Methods}
\subsection{Training Sets}
The training sets and reference computations of all molecules are based on the multi-reference configuration interaction method accounting for single and double excitations (MR-CISD) out of the active space of 6 electrons in 4 orbitals  with the double-zeta basis set aug-cc-pVDZ (augmented correlation consistent polarized valence double zeta) as implemented in Columbus.~\cite{Lischka2001PCCP} The molecules investigated in this study are the methylenimmonium cation (CH$_2$NH$_2^+$), ethylene (C$_2$H$_4$), aminomethylene (CHNH$_2$), methylenimine (CH$_2$NH), and C$_2$H$_5^+$. 
\paragraph{Training Sets}
The training set of the methylenimmonium cation, CH$_2$NH$_2^+$, forms the basis, as this training set already exists and can be taken from Ref.~\citenum{Westermayr2019CS}. It consists of 4,000 data points of three singlet states, which has been shown to cover the relevant configurational space visited after photo-excitation to the second excited singlet state, S$_2$. 

In order to compute an ample training set for ethylene in the most efficient way, the molecular geometries of the available CH$_2$NH$_2^+$ training set are used and the nitrogen atom is replaced by a carbon atom. 3,969 MR-CISD/aug-cc-pVDZ calculations are converged, with which the training set for ethylene is built. No optimizations of state minima or crossing points are carried out as this would lead to considerably higher computational effort. The same reference method as for the training set of CH$_2$NH$_2^+$ is used in order to allow for merging of the two training sets. Hence Rydberg states of ethylene are not described, which have also been neglected in some previous studies~\cite{Ben-Nun1998CPL,Granucci2001JCP,Quenneville2003JPCA,Angeli2009JCC,Tao2011JCP,Allison2012JCP,Hollas2018JCTC} and are considered to be less relevant in two-state photodynamics.~\cite{Granucci2001JCP,Mori2012JPCA}

As CH$_2$NH$_2^+$ is considered to be a three-state problem with a bright second excited singlet state and C$_2$H$_4$ is referred to as a two-state problem with a bright first excited singlet state, S$_1$,~\cite{Barbatti2006MP,Tapavicza2007PRL,Tavernelli2009JCP,Tavernelli2009JMS} these two molecules and their distinct photodynamics are considered to be a perfect testbed for the purpose of this study.

\subsection{Absorption Spectra and Electrostatic Potentials} Statistically significant results for the computation of UV/Visible absorption spectra can be obtained by sampling a lot of different molecular conformations. 
Here, the reference UV/Visible absorption spectra are obtained from excited-state calculations of 500 molecular conformations sampled from a Wigner distribution.~\cite{Wigner1932MM,Wigner1932PR}
The same method as for the training set generation is used for every molecule. Except for the equilibrium structure of CH$_2$NH$_2^+$ and C$_2$H$_4$, these 500 data points are not included in the training set. 
Alternatively, sampling could also be carried out with Born-Oppenheimer MD simulations, but Wigner sampling is considered to be superior for small molecules~\cite{Zobel2019PCCP} and is the standard procedure in SHARC.~\cite{sharc-md2} The calculated vertical excitations from every sampled conformation in combination with the corresponding oscillator strengths and a Gaussian broadening yield the UV/visible spectra. The width of the Gaussians are specified in table S1 in the SI.
In addition to the molecules, on which the ML models are trained on, the UV/Visible spectra of CH$_2$NH, CHNH$_2$, and C$_2$H$_5^+$ are computed from 500, 500, and 100 Wigner-sampled conformations. The molecular structures of these molecules are optimized at the MP2/TZVP level of theory using ORCA.~\cite{Neese2012WCMS} 

The electrostatic potentials are plotted with Jmol~\cite{jmol} and correspond to the energetically lowest lying conformation of each molecule. The Hirshfeld charges are obtained from MP2/TZVP calculations, while the Mulliken charges are available in Columbus, hence they are obtained from respective calculations with MR-CISD/aug-cc-pVDZ. 

\subsection{SchNarc}
As a deep learning model, SchNarc is used, which combines the continuous-filter convolutional-layer neural network SchNet~\cite{Schuett2018JCP,Schuett2019JCTC} for excited states and the MD program SHARC (Surface Hopping including ARbitrary Couplings).~\cite{sharc-md2,Richter2011JCTC,Mai2018WCMS} 
As the SchNarc model, originally developed for photodynamics, is described in details elsewhere,~\cite{Westermayr2020JPCL} thus we only shortly describe the technical details and timings of the computations.

\begin{figure*}[ht]
    \centering
    \includegraphics[scale=0.6]{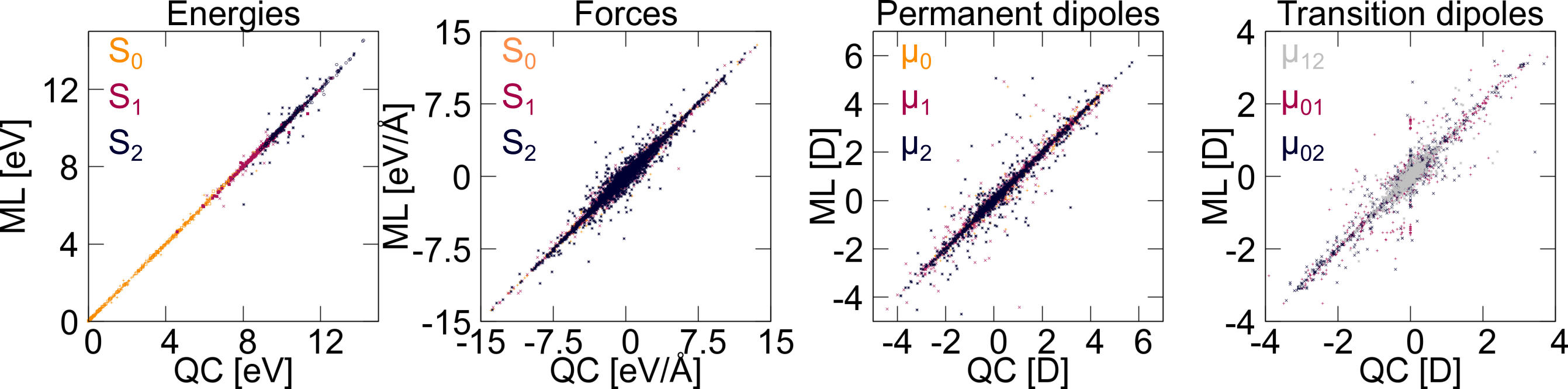}
    \caption{Scatter plots showing the reference energies, forces, permanent, and transition dipole moments plotted against the ML predictions of the model trained on CH$_2$NH$_2^+$ and C$_2$H$_4$ simultaneously.
    }
    \label{fig:scatter}
\end{figure*}

As ML is computationally efficient compared to quantum chemistry more conformations can be sampled and more trajectories can be initiated, while still being computationally less expensive. To this aim, 20,000 initial conditions are sampled from a Wigner distribution, from which the UV/Visible absorption spectra are computed using the oscillator strengths obtained from ML energy gaps and transition dipole moments in combination with Gaussian broadening. 
The computation of the three potential energies at 500 and 20,000 initially sampled molecular conformations takes about 9 sec (39 sec) and 6 min (26 min), respectively, on a GeForce GTX 1080 Ti GPU (Xeon E5-2650 v3 CPU) using the largest trained ML model. In contrast, 500 computations of three PESs with MR-CISD/aug-cc-pVDZ take about 17 hours on a Intel Xeon E5-2650 v3 CPU. 

SchNarc models are trained on 3,000 data points of CH$_2$NH$_2^+$ and C$_2$H$_4$ separately using 200 additional data points for validation during training and the remaining points for the test set. 5 hidden layers and 256 features to describe the atoms within a cut-off region of 5 {\AA} are used to generate the molecular descriptors. The model which is trained on both molecules takes 7,000 data points, 500 data points are used for validation and the rest is held back as a test set. The network architecture comprises 7 hidden layers and 512 features with a cut-off region of 6 {\AA}. The training of the single-molecule SchNarc model takes about 11 hours and of the model trained on both molecules with the larger network architecture about 15 hours on the aforementioned GPU. 

The trade-offs for each trained property along with the mean absolute error (MAE) obtained from all 3 states or all possible pairs of states on the test set of each model is given in Table~\ref{tab:error}. The scatter plots of the models are shown in Fig.~\ref{fig:scatter}. The largest errors can be estimated from the scatter plots. Especially in critical regions of the PESs, quantum chemical calculations are difficult to converge and can show artifacts and energy jumps in PESs,~\cite{Westermayr2019CS,Westermayr2020MLST} hence the scatter plots should be taken with care. The predicted dipole moments obtained with SchNarc are about a factor of 5 more accurate than our previously reported kernel ridge regression models~\cite{Westermayr2020MLST} and multi-layer feed-forward neural networks,~\cite{Westermayr2019CS,Westermayr2020MLST} which fit dipole moments in a direct way -- as single values with kernel ridge regression and as single elements put together in one vector with neural networks.

\begin{table}[ht]
    \centering
    \begin{tabular}{c|c|c}
         Model& MAE (RMSE) Energy [eV] & t$_E$  \\
         \hline
         CH$_2$NH$_2^+$& 0.047 (0.13) &1.0 \\
         C$_2$H$_4$ &0.11 (0.23) &1.0 \\
         Combined  &0.060 (0.15) &1.0 \\
         \hline \hline
 & MAE (RMSE) Forces [eV/{\AA}] & t$_F$\\
         \hline
         CH$_2$NH$_2^+$& 0.21 (0.49) &1.0 \\
         C$_2$H$_4$ & 0.32 (0.63) &1.0 \\
         Combined  &0.23 (0.52) &1.0 \\        
         \hline\hline 
         & MAE (RMSE) Dipoles [D] & t$_\mu$ \\
         \hline
         CH$_2$NH$_2^+$& 0.14 (0.44) &0.001 \\
         C$_2$H$_4$ & 0.19 (0.39)&0.1 \\
         Combined  & 0.13 (0.31) &0.2 \\
    \end{tabular}
    \caption{Trade-offs used to train energies, forces and dipole moments along with the mean absolute error (MAE) and root mean squared error (RMSE) on the test set for each property. Permanent and transition dipole moments are shown together as they are processed together with SchNarc. The mean over all states and pairs of states is shown. The respective scatter plots are given in Fig.~\ref{fig:scatter}.}
    \label{tab:error}
    \end{table}

\section{Results and discussion}

\subsection{UV/Visible Absorption Spectra}

\begin{figure*}[ht]
    \centering
    \includegraphics[scale=0.6]{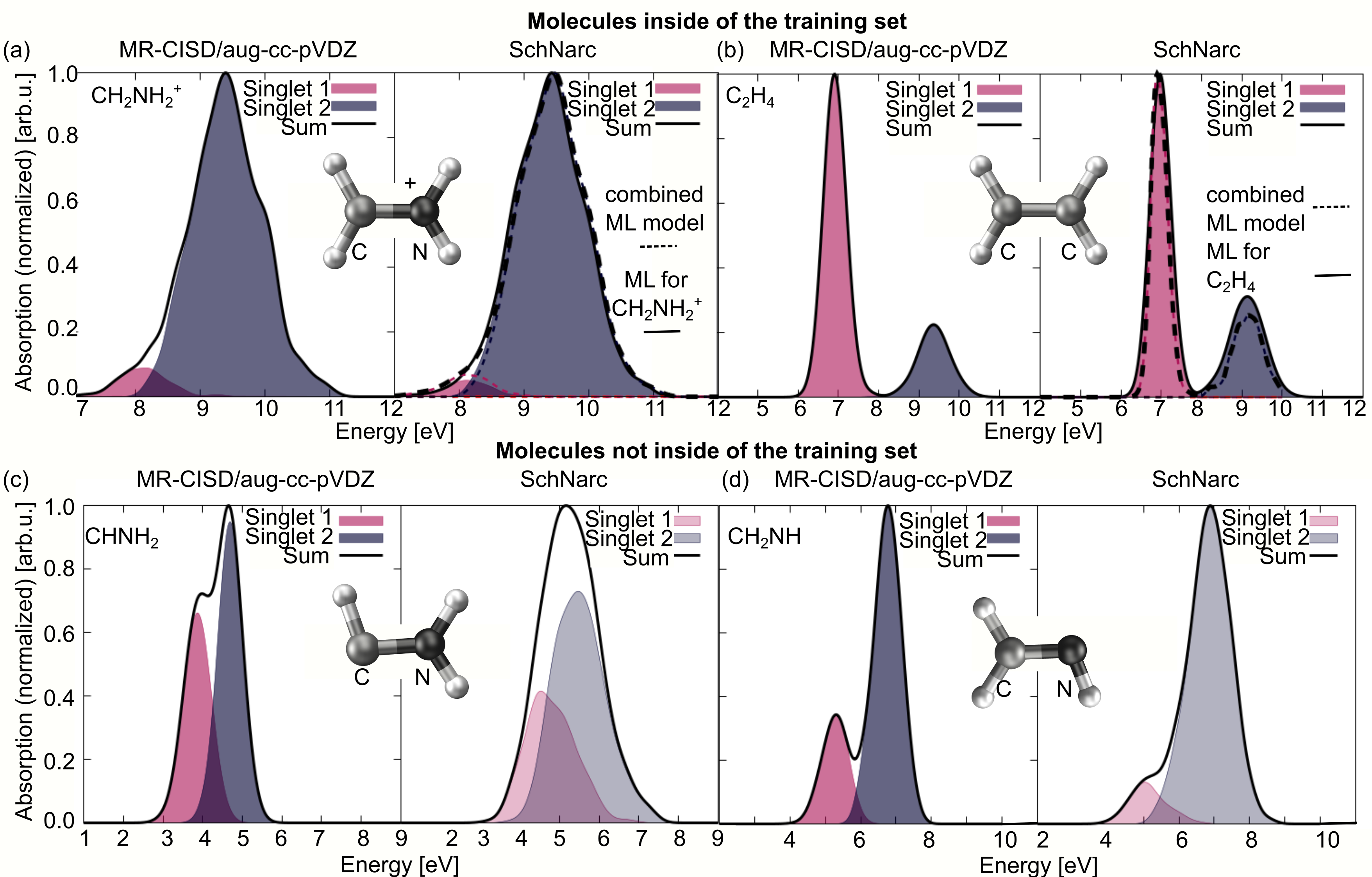}
    \caption{UV/Visible absorption spectra computed from 500 Wigner sampled conformations with MR-CISD/aug-cc-pVDZ on the left and from 20,000 Wigner sampled conformations with SchNarc on the right. The molecules (a) CH$_2$NH$_2^+$ and (b) C$_2$H$_4$ are included in the training set and the performance of the ML model trained on one (solid lines) and both molecules (dashed lines) are compared, while (c) CH$_2$NH and (d) CHNH$_2$ are not included in the training set and the ML model trained on both molecules is used for the prediction.}
    \label{fig:uv}
\end{figure*}

The computed UV/Visible absorption spectra are shown in Fig.~\ref{fig:uv} with the reference method on the left and the ML predictions on the right. 
Panels (a) and (b) illustrate the spectra of CH$_2$NH$_2^+$ and C$_2$H$_4$, which are both included in the training set. The filled spectrum with solid lines is obtained from SchNarc models trained solely on  CH$_2$NH$_2^+$ or C$_2$H$_4$ (i.e., a single-molecule model) and the dotted lines are obtained from the SchNarc model trained on the combined training set, which includes both molecules (i.e., a multi-molecule model). As it is visible, both models can be used to accurately predict the UV/Visible absorption spectra. Remarkably, the S$_2$ state is correctly predicted to be bright for CH$_2$NH$_2^+$ in panel (a), while the S$_1$ state is dark, while the inverse relation is predicted correctly in panel (b) for C$_2$H$_4$. The results indicate that although the transition dipole moments are completely different for the different electronic states in both molecules, SchNarc can accurately capture the absorption behaviour of both molecules. Remarkably, the model trained on both molecules is even slightly more accurate than the ML model trained solely on C$_2$H$_4$ for the prediciton of the UV/Visible absorption spectrum in panel (b). We did not expect such an outcome because force fields with increasing generality become usually less accurate for specific examples.

Due to the advantage of the atom-wise molecular descriptor, which enables a description of different molecules of different sizes, the transferability capabilities of SchNarc for the prediction of a manifold of PESs and transition dipole moments throughout chemical compound space is evaluated. To this aim, the UV/Visible spectra of CHNH$_2$ and CH$_2$NH are additionally computed, which are shown in panels (c) and (d), respectively. 
In order to make sure that these molecules are not included in the training set, an analysis of the maximum bond distances in the training set is carried out. According to unrelaxed dissociation scans of CH$_2$NH$_2^+$, the hydrogen atoms can be considered as dissociated at a bond length of about 2.5 {\AA}. No geometry inside of the training set has an N-H bond length larger than 2 {\AA}, and eight geometries have a C-H bond length larger than 2 {\AA}, where only one is larger than 2.5 {\AA}. The same is true for the training set of C$_2$H$_4$ with regard to the C-H bond length. Thus, it can be safely said that the assessment of the performance of SchNarc is not biased by an unusual large amount of dissociated configurations in the training set.

As it is clearly visible in panels (c) and (d), the energies of the S$_1$ and S$_2$ states are lower compared to the energies of CH$_2$NH$_2^+$ and C$_2$H$_4$ in panels (a) and (b). This trend is predicted correctly with SchNarc for both CH$_2$NH and CHNH$_2$. Also the bright and dark states are predicted qualitatively correct. In panel (d), the S$_1$ state is much darker than the S$_2$ state, whereas the S$_1$ state is brighter in panel (c). Although the spectra of the SchNarc models of the unknown molecules are broadened compared to the quantum chemical spectra, they can be used to obtain a qualitatively correct picture of the UV/Visible light absorption at almost no additional costs.

CH$_2$NH and CHNH$_2$ both contain one atom less than the molecules described in the training set. Thus, one might assume, that also the ML model trained solely on CH$_2$NH$2^+$ can be used to predict a qualitatively correct UV/Visible absorption spectra, as only atoms have to be removed. However, evaluation of the single-molecule models shows that this model cannot be used to capture the correct absorption behaviour and energy range of the two molecules not included in the training set. The performance of the ML model trained solely on CH$_2$NH$_2^+$ is even comparable to the ML model trained solely on C$_2$H$_4$, which would be expected to be at least worse. 

As already indicated, the molecular structures of the tested molecules, CH$_2$NH and CHNH$_2$, are similar to CH$_2$NH$_2^+$ and C$_2$H$_4$. 
In order to assess the performance of SchNarc for the computation of the UV/Visible absorption spectra of molecules with a different structure, the isoelectronic molecule C$_2$H$_5^+$, which contains one atom more, is additionally chosen.
\begin{figure}[ht]
    \centering
    \includegraphics[scale=0.3]{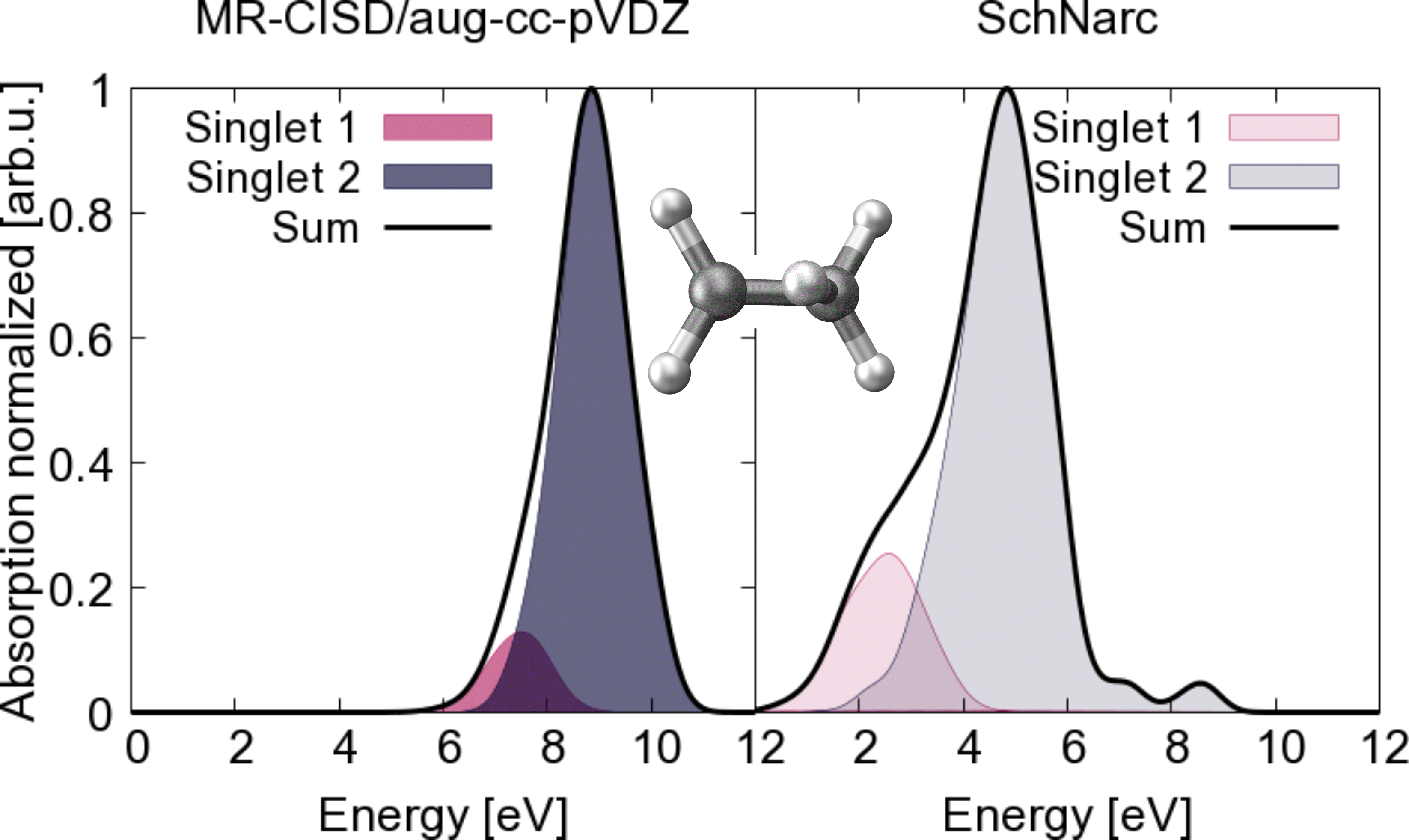}
    \caption{UV/Visible absorption spectrum of C$_2$H$_5^+$ computed with MR-CISD/aug-cc-pVDZ for two excited singlet states from 100 Wigner sampled conformations. 20,000 Wigner sampled geometries are used to obtain the spectrum on the left computed with SchNarc trained on CH$_2$NH$_2^+$ and C$_2$H$_4$.}
    \label{fig:uv_bad}
\end{figure}

Fig.~\ref{fig:uv_bad} shows the reference spectrum on the left and the ML-predicted spectrum on the right. The trained SchNarc models cannot be used to predict the UV/Visible absorption spectrum of C$_2$H$_5^+$. While the S$_1$ state is predicted to be dark and the S$_2$ state to be bright, which is in accordance to the reference spectrum, the energy range is off. Reasons can be the larger system size, due to the different shapes of the molecules, or due to both reasons. As three hydrogen atoms are bound to a carbon atom in C$_2$H$_5^+$, the structure of this molecule is completely different to the structures inside of the training set. 

The results shown here leave us to conclude that isoelectronic molecules with similar molecular structure can be predicted and that our ML models are to a certain extent transferable throughout chemical compound space also for excited-state PEss and properties thereof. It would be interesting to assess the transferability capacity of ML for the excited states when treating a larger number of molecules. Unfortunately, the high expenses and complexity of multi-reference quantum chemistry methods remain a clear bottleneck in this regard.

\subsection{Electrostatic potentials}
\begin{figure}[ht]
    \centering
    \includegraphics[scale=0.45]{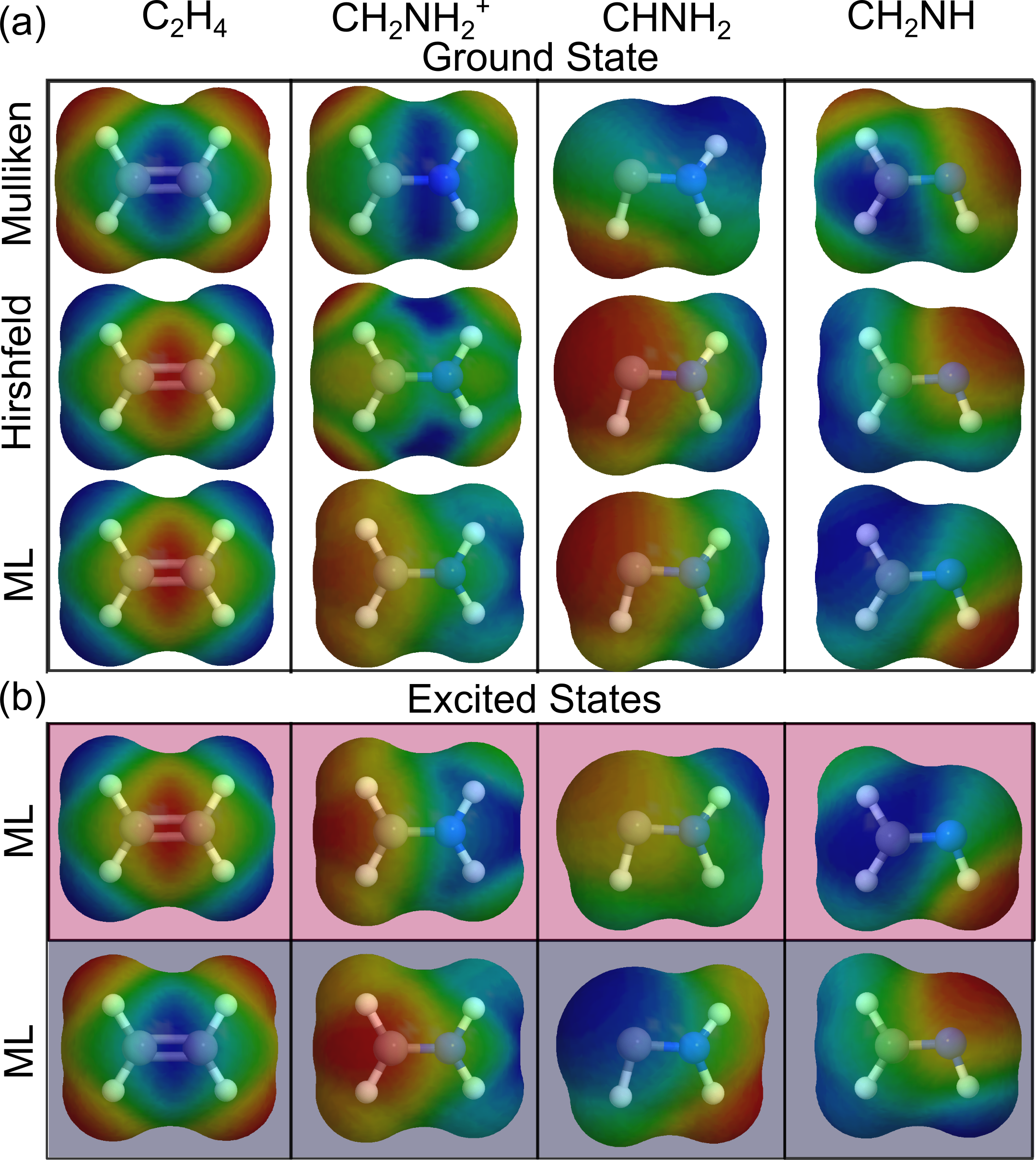}
    \caption{Electrostatic potentials of C$_2$H$_4$, CH$_2$NH$_2^+$, CHNH$_2$ and CHNH$_2$ obtained from Mulliken charges, Hirshfeld charges computed at MP2/TZVP level of theory, and latent charges of the ML models trained on the training set containing both C$_2$H$_4$ and CH$_2$NH$_2^+$ molecules for the (a) electronic ground state and (b) excited states -- the latter being only predicted with ML. Reddish colors indicate regions of negative charge, while blue refers to positive charges.}
    \label{fig:esp}
\end{figure}

The transition dipole moments and energies provide a measure of the quality of the molecular properties that are constructed from atomic contributions with SchNarc. As mentioned above, SchNarc also provides direct access to latent ground-state and excited-state partial charges based solely on the dipole moment data of the underlying electronic structure method.
In order to assess, whether the ML model provides meaningful partial charges, the electrostatic potentials obtained from SchNarc are compared to those obtained from Mulliken and Hirshfeld charges. Note that the latter are rarely implemented in quantum chemistry programs for excited states. The results are thus shown only for the electronic ground state in Fig.~\ref{fig:esp}(a).

The first and the second column show the molecules, which are included in the training set. Red colors indicate negative charges, while blue colors indicate positive charges. The electrostatic potentials in the first line are obtained from Mulliken charges. As it is visible, the Mulliken scheme shows that negative charges are located at hydrogen atoms and positive charges at the carbon atoms, which is in contrast to the Hirshfeld scheme given in the second line and also in contrast to chemical intuition. The electrostatic potentials obtained from the ML model trained on both molecules is shown in the third line. A similar charge distribution is obtained for ML models trained on a single molecule (see Fig. S2 in the SI). The partial charges obtained from SchNarc are in good agreement with the Hirshfeld charges. Similar agreement, at least qualitatively, can be obtained for CH$_2$NH$_2^+$.

As the charge distribution of the electronic ground state is in qualitatively good agreement to the Hirshfeld partitioning scheme, the redistribution for the excited states can be analyzed. In case of C$_2$H$_4$ in the first column of panel (b), the negative and positive charges do not redistribute considerably in case of the S$_1$ state, but the distribution is inverted for the S$_2$ state.  The positive charge is then located between the carbon atoms. For CH$_2$NH$_2^+$, the positive charge is located at the far end of the nitrogen side of the molecule for the ground state. In the S$_1$ state, the positive charge is still located at the nitrogen but closer to the center of the molecule. In the S$_2$ state, the distribution is similar to the ground state. These distributions give rise to dipole moments, which perfectly agree with the reference calculations (QC/ML: 1.5 a.u. (S$_0$), 1.2 a.u. (S$_1$), 1.5 a.u. (S$_2$); the vectors all point from C towards N).  

In addition to the molecules included in the training set, the transferability of SchNarc to predict electrostatic potentials is tested too. Although the ML model has never been trained on CHNH$_2$ or CH$_2$NH, the ground-state electrostatic potentials agree arguably better with the Hirshfeld distributions than the Mulliken ones. This is especially true for CHNH$_2$. Comparing the S$_0$ distribution with the one from S$_2$, an inversion of the charge locations is visible, which is also present in C$_2$H$_4$ but not in CH$_2$NH$_2^+$. 

The last column illustrates the electrostatic potentials of CH$_2$NH, where the negative charge is located at the nitrogen atom according to the Hirshfeld partitioning but rather at the adjacent hydrogen according to ML. 

All these results indicate that the charge distributions obtained with SchNarc can be used to obtain electrostatic potentials of molecules included in the training set and that transferability is possible also for latent partial charges, at least for isoelectronic molecules. 

\section{summary and Outlook}

In this work, the SchNarc deep learning approach for photodynamics is extended to describe permanent and transition dipole moments in a rotationally covariant manner and for an arbitrary number of electronic states. The dipole moment vectors can be trained in one ML model in addition to the ground-state energies and forces as well as a manifold of excited-state energies and forces. SchNarc can be used to accurately predict UV/Visible absorption spectra and the latent partial charges can be used to assess the charge distribution via electrostatic potentials of molecules.
As SchNarc is trained not only on the ground state, but also on the excited states, the charge distribution for the excited states can be assessed. As the partial charges for the ground state are in qualitatively good agreement to the Hirshfeld charges and also the excited-state molecular dipole moments agree between ML and the reference, we consider the charges to be equally accurate also for the excited states. The latent partial charges are based on highly accurate quantum chemistry and provide direct access to the charge distribution after light excitation. 

Due to the atom-wise tailored descriptor,
many different molecules can be described in one model, which contain different numbers of atoms. At least when isoelectronic, similarly structured molecules are treated, transferability is confirmed for UV/Visible absorption spectra and partial charges. These properties can be computed with our ML approach at least qualitatively at almost no additional costs. Remarkably, the ML model can treat charged species on the same footing as neutral species.

Especially interesting would be to assess the improvement one can achieve by including many more molecules than just two isoelectronic ones. At the current stage of research, the high complexity and costs of accurate multi-reference quantum chemical methods hampers an ample assessment of the transferability in the excited states. Nevertheless, the trend clearly shows that ML models trained on more molecules are superior to ML models trained on single molecules, even if these molecules exhibit a completely different photochemistry and overall charge. 

\section*{Data availability}
The training set for CH$_2$NH$_2^+$ is published with Ref.~\citenum{Westermayr2019CS} and the training set of C$_2$H$_4$ will be made available as supporting information in the same format as the previous training set, i.e., the one used by SHARC.~\cite{sharc-md2} 
The SchNarc model is updated and freely available at \href{https://github.com/schnarc/schnarc}{https://github.com/schnarc/schnarc}.
\begin{acknowledgments}
This  work was financially supported by the AustrianScience Fund, W 1232 (MolTag) and the uni:docs program of the University of Vienna (J.W.). The computational results presented have been achieved in part using the Vienna Scientific Cluster (VSC). P. M. thanks the University of Vienna for continuous support, also in the frame of the research platform ViRAPID. 
The authors thank Michael Gastegger for helpful discussions regarding the extension of the charge model of Ref.~\citenum{Gastegger2017CS} for SchNarc and concerning the latent partial charges.
\end{acknowledgments}

%

\renewcommand{\thefigure}{S\arabic{figure}} 
\renewcommand{\thetable}{S\arabic{table}} 
\renewcommand{\thesection}{S\arabic{section}}
\setcounter{equation}{0}
\setcounter{figure}{0}
\setcounter{table}{0}
\setcounter{section}{0}

\clearpage 
\onecolumngrid
\begin{center}
\Huge{\textsf{\textbf{Supporting Information}}}
\end{center}
\twocolumngrid
\section{UV/Visible Absorption Spectra}

\begin{figure}[b]
    \centering
    \includegraphics[scale=0.48]{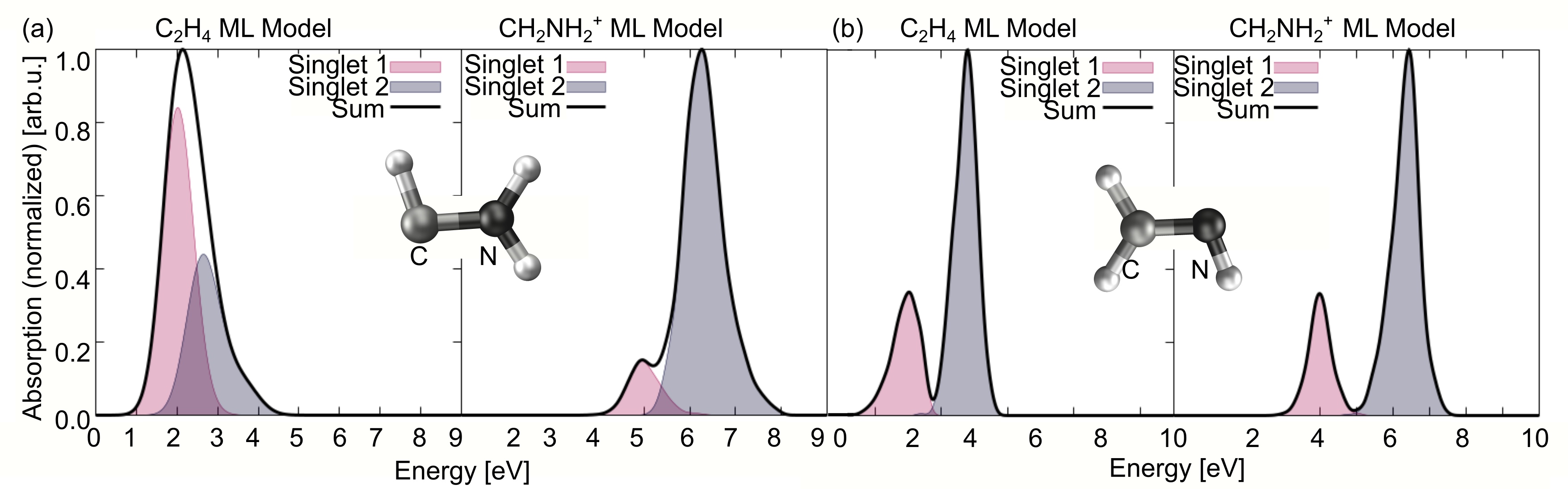}
    \caption{UV spectra of (a) CH$_2$NH and (b) CHNH$_2$ predicted with the ML models trained solely on C$_2$H$_4$ (left plots) and CH$_2$NH$_2^+$ (right plots). The minimum and maximum energy was selected according to Fig. 2 and was extended where necessary in order to enable better comparison between the spectra.  A full width at half maximum of 0.3 eV for all spectra, expect for panel (a) using the C$_2$H$_4$-ML model, where the width is set  0.75 eV.}
    \label{fig:si_uv}
\end{figure}

UV/Visible absorption spectra are computed from energy differences and transition dipole moments applying Gaussian broadening.
Dependent on the number of sampled conformations a full width at half maximum (FWHM) for the Gaussian convolution of 0.5-1.0 eV is used for the quantum chemical spectra and of 0.3-0.5 eV for the ML-predicted spectra in Fig. 2 in the main text and Fig.~\ref{fig:si_uv}. The reason is to avoid an unphysical fine structure of the spectra, resembling vibrational quantum levels although only electronic degrees of freedom are quantized in the employed approach. The width of the Gaussian used is specified in Table~\ref{tab:fwhm}. Noticeably, the sampling of even more molecular conformations can reduce the FWHM, which has been shown recently to be possible with ML,~\cite{Xue2020chemrxiv} but is not the main purpose of this study.

\begin{table}[ht]
    \centering
    \begin{tabular}{c|ccccc}
         Method &C$_2$H$_4$ &CH$_2$NH$_2^+$ &CHNH$_2$ &CH$_2$NH &C$_2$H$_5^+$ \\
         \hline \hline
         QC& 0.5 &0.5&0.5&0.5&0.95\\
         ML-1 &0.3 &&0.3 &0.5&\\   
         ML-2&&0.3&0.3&0.3\\
         ML-12 &0.3&0.3&0.4&0.4&0.75\\
    \end{tabular}
    \caption{The used FWHM for the spectra computed with the quantum chemistry reference method MR-CISD/aug-cc-pVDZ (abbreviated as QC) using 500 molecular configurations for C$_2$H$_4$, CH$_2$NH$_2^+$, CHNH$_2$, and CH$_2$NH, and 100 molecular configurations for C$_2$H$_5^+$. ML-1, ML-2, and ML-3 denote the ML models trained on C$_2$H$_4$, CH$_2$NH$_2^+$, and both molecules, respectively. UV spectra are computed from 20,000 molecular configuration of each molecule.}
    \label{tab:fwhm}
\end{table}

The performance of the ML models trained on only one molecule, i.e., C$_2$H$_4$ (left plots) and CH$_2$NH$_2^+$ (right plots) separately, for the computation of UV/Visible spectra of the molecules  CHNH$_2$ and CH$_2$NH are compared in Fig.~\ref{fig:si_uv}(a) and Fig.~\ref{fig:si_uv}(b), respectively. As it is visible, the ML model trained on CH$_2$NH$_2^+$ predicts for both C$_2$H$_4$ and CH$_2$NH$_2^+$ the first excited singlet state to be darker than the second excited singlet state, which is also the case for CH$_2$NH$_2^+$, whose behaviour the ML model has learned. The spectrum of CHNH$_2$ is not comparable to the the reference spectrum shown in the main text in Fig. 2(b) at all. For CH$_2$NH, the resulting curves agree qualitatively with the reference spectrum, but the energy gap between the two absorption peaks is larger. The two peaks slightly overlap in the reference spectrum.  

In contrast, the ML model trained solely on C$_2$H$_4$ predicts the first excited singlet state to be brighter for CHNH$_2$ (panel (a) left plot), but the opposite behaviour for CHNH$_2$ (panel (b) left plot), which is comparable to the reference spectrum. However, the energy range is not comparable to the reference. 

The results here show that an ML model solely trained on one molecular species is not transferable, even though the molecule to be predicted contains a subset the same atoms (arranged in the same way). The ML models trained on both molecules discussed in the main text, however, show much better transferability, although the two molecules contained in the training set exhibit a different photochemistry. Our assumption that the ML model gets worse with each additional molecule in the training set is refuted. 

\section{Electrostatic Potentials}
For the training of dipole moment vectors, the simplest phase-less loss function is used, which is computationally efficient compared to more accurate loss functions reported in Ref.~\citenum{Westermayr2020JPCL}, which are necessary e.g. for photodynamics simulations based on couplings. Here, the minimum function for fitting the permanent dipole moments and transition dipole moments suffices, but as a consequence, the trained properties are only defined up to an arbitrary sign. While this does not influence the transition dipole moments, the signs of the permanent dipole moment vectors for each electronic state have to be adjusted with respect to a reference geometry, e.g., the ground-state equilibrium geometry.
\onecolumngrid

\twocolumngrid

If e.g. reaction scans are executed subsequently, the assigned sign has to be considered in order to obtain the correct direction of the permanent dipole moment vectors along the reaction path. The signs only have to be adjusted for one molecular geometry as the ML outputs are smooth functions by definition.~\cite{Westermayr2020JPCL} The manual assignment can be circumvented by applying the more accurate phase-less loss function (equations 3 and 4 in Ref.~\citenum{Westermayr2020JPCL}).  Nevertheless, the comparison of the signs for one molecular geometry is rather inexpensive compared to a much longer training procedure.

As the number of electrons is not encoded in the descriptor and the overall charge of the molecule is not known, the atomic partial charges have to be scaled in order to resemble the correct molecular charges when using latent partial charges for electrostatic potentials for example. The scaled charges of atom a for a given electronic state i, $\tilde{q}_{i,a}$:
\begin{equation}
    \tilde{q}_{i,a}=q_{i,a}-\frac{1}{N_a}\left( \sum_a^{N_a}q_{i,a}-Q\right)
\end{equation}
with N$_a$ being the number of atoms in a molecule and Q the charge of the molecule. 

\begin{figure}[ht]
    \centering
    \includegraphics[scale=0.5]{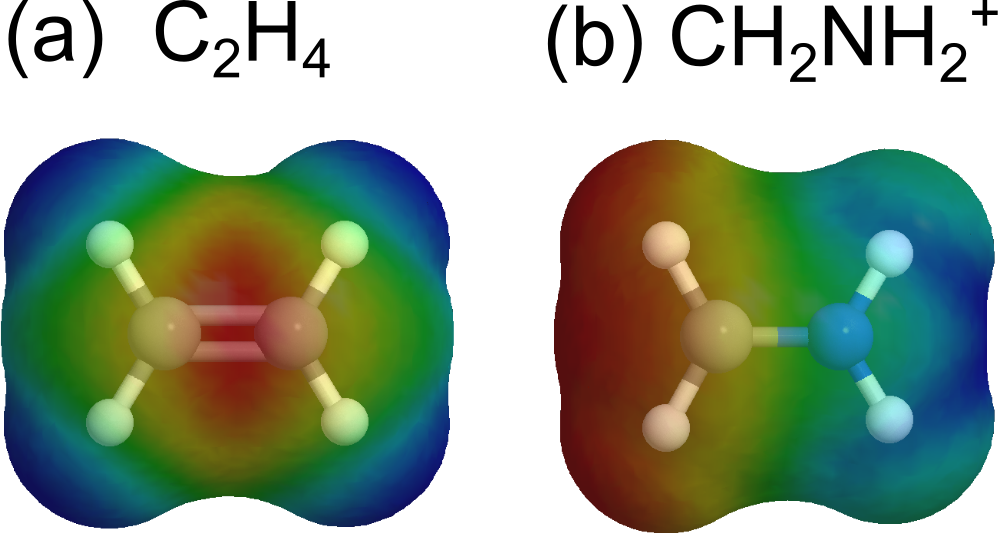}
    \caption{Electrostatic potentials predicted for (a) C$_2$H$_4$ and (b) CH$_2$NH$_2^+$ using the latent charges of the respective single-molecule ML model. Reddish colors indicate regions of negative charge, while blue refers to positive charges.}
    \label{fig:si_esp}
\end{figure}

Electrostatic potentials of C$_2$H$_4$ and CH$_2$NH$_2^+$ computed with the single-molecule ML models fitted on C$_2$H$_4$ and CH$_2$NH$_2^+$, respectively, are given in Fig.~\ref{fig:si_esp}. Comparison to electrostatic potentials obtained with Mulliken and Hirshfeld charges in Fig. 4 of the main text demonstrates, that C$_2$H$_4$ (panel (a)) 
is similar to Hirshfeld charges and thus also in excellent agreement to the model trained on both molecules.
The electrostatic potential computed with CH$_2$NH$_2^+$ in panel (b) for the CH$_2$NH$_2^+$ is also comparable to the ML model trained on both molecules.

\end{document}